\documentclass[superscriptaddress,pra,twocolumn]{revtex4}

\usepackage{graphicx}
\usepackage{hyperref}

\begin{document}

\title{Experimental investigation of amplitude and phase quantum correlations in a type II OPO above threshold: from the non-degenerate to the degenerate operation}

\author{Julien Laurat}\affiliation{Laboratoire
Kastler Brossel, Case 74, Universit{\'e} P. et M. Curie, 4 Place
Jussieu, 75252 Paris cedex 05, France}

\author{Thomas Coudreau}\email{coudreau@spectro.jussieu.fr}\affiliation{Laboratoire
Kastler Brossel, Case 74, Universit{\'e} P. et M. Curie, 4 Place
Jussieu, 75252 Paris cedex 05, France}\affiliation{Laboratoire
Mat{\'e}riaux et Ph{\'e}nom{\`e}nes Quantiques, Case 7021, Universit{\'e} D.
Diderot, 2 Place Jussieu, 75251 Paris cedex 05, France}

\author{Laurent Longchambon}\affiliation{Laboratoire
Kastler Brossel, Case 74, Universit{\'e} P. et M. Curie, 4 Place
Jussieu, 75252 Paris cedex 05, France}

\author{Claude Fabre} \affiliation{Laboratoire
Kastler Brossel, Case 74, Universit{\'e} P. et M. Curie, 4 Place
Jussieu, 75252 Paris cedex 05, France}

\date{\today}

\begin{abstract}
We describe a very stable type II optical parametric oscillator
operated above threshold which provides 9.7 $\pm$ 0.5 dB (89\%) of
quantum noise reduction on the intensity difference of the signal
and idler modes. We also report the first experimental study by
homodyne detection of the generated bright two-mode state in the
case of frequency degenerate operation obtained by introducing a
birefringent plate inside the optical cavity.
\end{abstract}

\maketitle

Type II optical parametric oscillators are well-known to generate
above threshold highly quantum correlated bright twin beams.
Intensity correlations were experimentally observed several years
ago and applied to measurements of weak physical effects
\cite{Heidmann87,Schwob97,Gao98}. Phase anti-correlations are also
theoretically predicted \cite{Reynaud}: a type II OPO above
threshold could be thus, in principle, a very efficient source of
bright EPR beams, which can be used in continuous variable quantum
information protocols such as cryptography, teleportation, secret
sharing or optical-atomic interfacing \cite{CV}. However, the non
frequency-degenerate operation makes difficult the study of the
phase properties by usual homodyne detection techniques
\cite{brazil}. Frequency degeneracy occurs only accidentally since
it corresponds to a single point in the experimental parameter
space. Actually, up to now, no direct evidence of
anti-correlations has been observed. The generation of EPR beams
with type II OPO has been thus restricted to the below threshold
regime where it behaves as a passive amplifier
\cite{Ou92,Zhang00,Laurat04b}, unlike above threshold where it is
an active oscillator choosing its working point.

In 1998, while working on optical frequency divider, E.J. Mason
and N.C. Wong proposed an elegant way to achieve frequency
degenerate operation above threshold \cite{Mason98,Fabre99}: a
birefringent plate inside the optical cavity and making an angle
with the axis of the non-linear crystal induces a linear coupling
between the signal and idler and results in a locking phenomenon,
which is well-known for coupled mechanical or electrical
oscillators \cite{Pikovsky}. In this original device called
"self-phase-locked" OPO, we have shown theoretically that quantum
correlations are preserved for small angles of the plate and that
the system produces non separable states in a wide range of
parameters \cite{longcham2}. We describe here an improvement of
the intensity correlations in a "standard" OPO and then report the
first experimental demonstration of homodyne detection operated on
the bright two-mode state. Our results are finally interpreted as
polarization squeezing.

\begin{figure}[b]
\includegraphics[width=0.95\columnwidth]{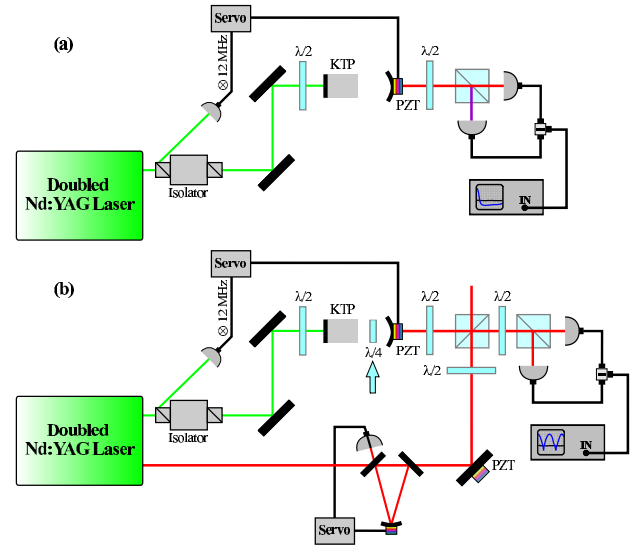}
\caption{A doubled Nd:YAG laser pumps above threshold a type II
OPO (a) without or (b) with a $\lambda/4$ plate inside the cavity.
In (a), intensity correlations are directly measured by a balanced
detection scheme. In (b), the frequency-degenerate operation opens
the possibility to implement an homodyne detection. The infrared
output of the laser is used as local oscillator after
filtering.}\label{setup}
\end{figure}

The experimental setup is shown in Fig. \ref{setup}. A
continuous-wave frequency-doubled Nd:YAG laser ("Diabolo",
Innolight GmbH) pumps a triply resonant OPO above threshold, made
of a semi-monolithic linear cavity: in order to improve the
mechanical stability and reduce the reflections losses, the input
flat mirror is directly coated on one face of the 10mm-long KTP
crystal. The intensity reflection coefficients for the input
coupler are 95\% for the pump at 532 nm and almost 100\% for the
signal and idler beams at 1064 nm. The output mirror, with a
radius of curvature of 38 mm, is highly reflective for the pump
and its transmission coefficient $T$ can be chosen to be 5 or
10\%. With $T=5\%$, at exact triple resonance, the oscillation
threshold is less than 15 mW. The OPO length is actively locked on
the pump resonance by the Pound-Drever-Hall technique: a remaining
12 MHz modulation present in the pump laser is detected by
reflection and the error signal is sent to a home-made
proportional-integral controller. In order to stabilize the OPO
output infrared intensity, the temperature of the crystal is
servo-locked at the requested temperature within a mK. In spite of
the triple resonance which generally makes OPOs much more
sensitive to disturbances, these controls enable a long-term
stability: the OPO operates stably during more than one hour
without mode-hopping. Depending on the presence of the plate, the
output state is characterized by different techniques.

\begin{figure}
\includegraphics[width=0.85\columnwidth]{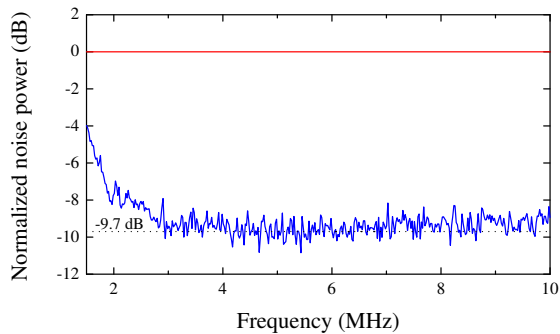}
\caption{Normalized noise power of the intensity difference of the
signal and idler as a function of the frequency, after correction
of the electronic noise. A reduction of 9.7 $\pm$ 0.5 dB is
reached around 5 MHz.}\label{TwinBeams}
\end{figure}
Without the plate, intensity correlations are directly measured by
a balanced detection scheme (Fig. \ref{setup}(a)). The signal and
idler orthogonally polarized beams, with different frequencies,
are separated on a polarizing beam splitter and detected on a pair
of high quantum efficiency InGaAs photodiodes (Epitaxx ETX300). A
half-wave plate is inserted before the polarizing beam splitter.
When the polarization of the twin beams is turned by 45$^{\circ}$
with respect to its axes, it behaves as a 50-50 usual beam
splitter, which allows to measure the shot noise level. With a
transmission $T=10\%$ for the output mirror, we obtained a noise
reduction of $9.7 \pm 0.5$ dB (89\%) around 5 MHz (Fig.
\ref{TwinBeams}). To the best of our knowledge, this noise
reduction is the strongest reported to date in the experimental
quantum optics field.

Intensity correlations can be measured even with non-frequency
degenerate beams but the measurement of phase anti-correlations is
much easier with degenerate beams and the use of a local
oscillator. To achieve the frequency-degenerate operation, a
birefringent plate is introduced inside the OPO cavity
\cite{Mason98,Fabre99}. This plate is chosen to be exactly
$\lambda/4$ at 1064 nm and almost $\lambda$ at 532 nm pump
wavelength. For a well-defined range of parameters -- reached by
the adjustment of both the crystal temperature and the frequency
of the pump laser -- a frequency locking phenomenon occurs and can
be maintained during more than hour. Degenerate operation is
confirmed by interference of the generated beams with a local
oscillator or, more directly, by the fact that the generated mode
has now a fixed polarization. At the minimum threshold point, the
generated state is linearly polarized at +45$^{\circ}$. An
ellipticity around 2\% has been measured.

The theoretical quantum properties of the device have been studied
by L. Longchambon \textit{et al.} \cite{longcham2}: for a small
angle of the plate, with respect to the transmission of the output
mirror, quantum correlations and anti-correlations are preserved.
Instead of measuring correlations and anti-correlations, it is
equivalent to measure the noise spectrum of the $\pm$45$^{\circ}$
polarized modes, $A_+$ and $A_-$. These orthogonally polarized
modes have squeezed fluctuations: for instance the squeezing of
$A_+$ is given by the phase anticorrelations. It should be
stressed that this characterization method is strictly equivalent
to correlation measurements: the inseparability criterion
\cite{Duan00} is defined as the half sum of these two noise
reductions. Let us underline, due to the defined phase relation,
$A_+$ is a bright mode, which corresponds to the mean field, and
$A_-$ has a zero mean value.

\begin{figure}
\includegraphics[width=0.8\columnwidth]{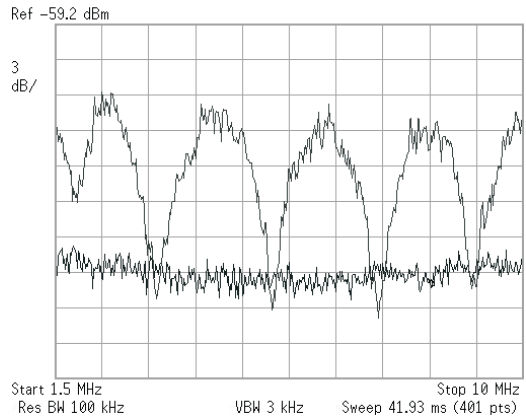}
\caption{Noise power of the mode $A_-$ while scanning
simultaneously the phase of the local oscillator and the noise
frequency between 1.5 and 10 MHz. The lower trace gives the shot
noise level.}\label{Amoins}
\end{figure}

The homodyne detection scheme is depicted on Fig. \ref{setup}(b).
The coherent 1064 nm output is used as a local oscillator after
filtering by a triangular 45cm-long cavity with a finesse of 3000.
This cavity is locked on the maximum of transmission by the
tilt-locking technique \cite{Shaddock99} and 80\% of transmission
is obtained. The fringe visibility reaches 0.97. The shot noise
level is obtained by blocking the output of the OPO. For the zero
mean value mode, $A_-$, this procedure directly gives the shot
noise level. For the bright mode, the incoming power is taken
equal to the power of the local oscillator: the shot noise level
is thus 3 dB higher than the measured noise.

Figure \ref{Amoins} gives the noise power of the mode $A_-$ while
scanning the local oscillator phase, for a transmission $T=5\%$
and a plate angle of 0.1$^{\circ}$. The noise frequency is also
scanned in order to give both the noise reduction and its
frequency dependance. As expected, more than 3 dB of squeezing is
obtained around a few MHz. One can also make a measurement as a
function of the local oscillator phase for a given noise
frequency: in this case we have measured at 3.5 MHz a value of
squeezing of 4.5 dB. This strong reduction on the mode $A_-$
confirms the quantum correlations of the signal and idler modes.

\begin{figure}
\includegraphics[width=0.8\columnwidth]{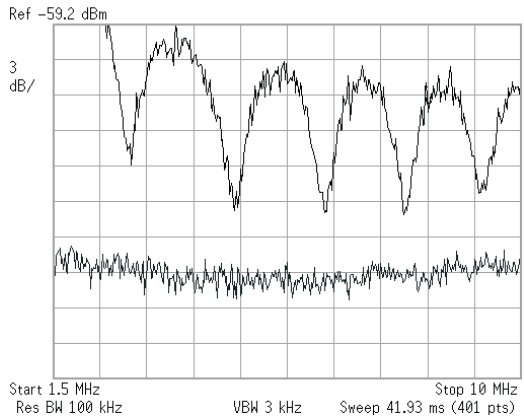}
\caption{Noise power of the mode $A_+$ while scanning the phase of
the local oscillator, for a noise frequency between 1.5 and 10
MHz. The shot noise level is given by the lower trace plus 3 dB.
}\label{Aplus}
\end{figure}

Figure \ref{Aplus} shows the noise power of the mode $A_+$ in the
same condition. A similar amount of noise reduction is expected.
However, a slight excess noise of 3 dB is measured for the minimal
noise quadrature: the anticorrelations are thus degraded, probably
by external noise sources.

Thus, despite this slight excess noise which prevents from
reaching EPR correlations, the generated state is squeezed on the
polarization orthogonal to the mean field: $A_+$ is the main mode
and $A_-$ the squeezed vacuum one. This condition is required to
obtain a so-called "polarization squeezed" state
\cite{Korolkova,Bowen,Josse}. 4.5 dB of polarization squeezing has
been thus generated by our original self-phase-locked OPO. This is
the first experimental demonstration of polarization squeezing
with an OPO above threshold. Such states have recently raised
great interest, in particular because of the possibility to map
quantum polarization state of light onto an atomic ensemble
\cite{Hald}.

In conclusion, we have built a compact and very stable type II
triply resonant OPO and explored this device above threshold, in
different regimes. Thanks to a great stability, the strongest
quantum noise reduction to date has been obtained. By adding a
plate inside the optical cavity, the frequency degenerate
operation then has been reached and has permitted the experimental
demonstration of homodyne detection operated above threshold. This
result opens a very promising way to the direct generation of
intense entangled beams and offers a new and simple method to
achieve strong polarization squeezing.

\begin{acknowledgments}
Laboratoire Kastler-Brossel, of the Ecole Normale Sup\'{e}rieure
and the Universit\'{e} Pierre et Marie Curie, is associated with
the Centre National de la Recherche Scientifique (UMR 8552). This
work has been supported by the European Commission project QUICOV
(IST-1999-13071) and ACI Photonique (Minist\`ere de la Recherche).
\end{acknowledgments}

\end{document}